\documentclass[aps,showkeys,showpacs,nosuperscriptaddress,twocolumn]{revtex4}

\usepackage{psfrag}
\usepackage{graphicx}
\usepackage{amsmath}
\usepackage{hyperref}
\usepackage{color}

\begin{document}

\title{Opinion formation in an open system and the spiral of silence}

\author{Przemys{\l}aw~Gawro\'nski}
\email{gawron@newton.ftj.agh.edu.pl}
\affiliation{
AGH University of Science and Technology, Faculty of Physics and Applied Computer Science, al. Mickiewicza 30, PL-30059 Krakow, Poland 
}

\author{Maria~Nawojczyk}
\email{maria@list.home.pl}
\affiliation{
AGH University of Science and Technology, Faculty of Humanities, al. Mickiewicza 30, PL-30059 Krakow, Poland 
}

\author{Krzysztof~Ku{\l}akowski}
\email{kulakowski@fis.agh.edu.pl}
\affiliation{
AGH University of Science and Technology, Faculty of Physics and Applied Computer Science, al. Mickiewicza 30, 30059 Krakow, Poland 
}

\date{\today}

\begin{abstract}
A new model is formulated of the sociological effect of the spiral of silence, introduced by Elisabeth Noelle-Neumann in 1974. The probability that a new opinion is 
openly expressed decreases with the difference between this new opinion and the perceived opinion of the majority. We also assume that the system
is open, i.e. some people enter and some leave during the process of the opinion formation. An influence of a leader is simulated by a comparison of two runs of the 
simulation, where the leader has different opinion in each run. The difference of the mean expressed opinions in these two runs persists long after the leader's leave. 
\end{abstract}



\maketitle

%% ###########################################################################
\section{Introduction}
\label{S1}
%% ###########################################################################

Opinion dynamics is often explored by physicists who enter to social sciences \cite{wc3,gc3,ds,szw,cfl,mgk}. There, they find applications for ideas and 
tools established in statistical physics, as Monte Carlo simulations \cite{moo}, mean field theory and phase transitions \cite{cont}, contact processes \cite{dak}, Ising model \cite{sta}, Master equations \cite{weii}, graphs \cite{sn,mjk}, synergetics \cite{wei}, self-organized criticality \cite{tur,ggb}, deterministic chaos \cite{kiel}, and so on. Being also involved in this field, yet we feel that more than often, the ways of thinking and formulating research questions by physicists and by social scientists do not meet. For a (socio)physicist it makes sense, then, to look for a problem formulated within sociology. Whilst our physics-oriented methods are our fate, at least the subject could be a common ground. Within the vast area of opinion dynamics research, here we are going to contribute to the theory of the spiral of silence (TSS) \cite{enn} by means of a new computational model. \\

The idea to formulate TSS mathematically is by no means new. An early contribution has been proposed by Granovetter and Soong in 1988 \cite{gso}, with using a set of nonlinear difference equations on the population dynamics; yet the concept of minority, basic for TSS, is not preserved there. In a parallel paper by Krassa \cite{krss}, the problem has been expressed in terms of a weighted social network of interacting agents. There, separate variables have been used to denote a significance of opinion of $i$-th for $j$-th agent, provided that $i$ did express his opinion on a given issue. Also, a set of thresholds have been assigned to particular agents, with their individual activity depending on the relation of an observed support for a given issue to the value of the threshold. Yet the model became so complex that the paper \cite{krss} does not contain any calculations. In a review paper \cite{25} published twelve years later by communication scholars, both these approaches \cite{gso,krss} remained unnoticed.
In a recent book \cite{book} devoted to TSS, the same omission is found. In subsequent simulations, either the Krassa theory has been applied \cite{bart,gly}, or a set of linear difference equations has been used \cite{wng}. In the latter case, the set of individualized variables denoted the willingness of an $i$-th agent to express at discrete time; with the rule of evolution dependent on the influence of media and the climate of the reference group of the agent. The idea of a collective action triggered by exceeding individual thresholds - a kind of domino effect - has been used also by Chwe \cite{chw}, without a direct relation to TSS.\\

Our goal here is to include the fact that the group content changes in time; some people enter, some people leave. Following the terminology of statistical physics, we dare to speak about open systems. The time scale of the variation of the group content depends on the group kind where opinions are formed, from tens of minutes in a queue or in a bar, to years in a school and tens of years in a multigenerational family, in a branch of science or in currents of a national culture. To keep our considerations within reasonable frames, we concentrate on TSS, where - up to our knowledge - the effect has not been discussed yet. In the literature mentioned above, the key issue is how the spiral of silence starts and develops. On the contrary, our main result is related to the question, how the effect is forgotten. We expect that the variation of the group content plays a major role here. We believe that the effect should be of relevance also for more general consideration of opinion dynamics. We are not aware of any 
approaches of this kind. In the formulation by Galam \cite{sga} which seems to be most close to ours, opinions of groups evolve in subsequent time steps, and group members are shuffled at the end of each step; yet, the content of the population as a whole is fixed and the mixing is just a mode of interaction between groups.\\

Our next section is devoted to the model formulation. Section III contains the numerical results on how long a leader is remembered after he/she leaves. In section IV we provide a simple argument that the term 'spiral' is justified in our model. Last section is reserved for final remarks.

%% ###########################################################################
\section{The model}
\label{S2}
%% ###########################################################################

According to the literature \cite{ennn}, TSS can be formulated in five hypotheses:\\

H1. Society threatens deviant individuals with isolation.\\

H2. Individuals fear isolation.\\

H3. This fear of isolation makes individuals scan continuously the opinion climate.\\

H4. The perceived climate of opinion influences their behaviour in public, specially in speaking out.\\

H5. (...) Those who think their opinion is the majority tend to speak out; otherwise they remain silent. Thus, the majority opinion seems to be more supported than it is in reality; 
the minority one seems to be less supported than it is. (...) At last stage, the minority opinion reduces to a small core group or disappears.\\

Following these hypotheses we keep or formulation within the agent-based modelling frames; for a concise review see \cite{gil}. As it was formulated in \cite{din}, an agent is 'an entity with goals'. In our case, the goal of an agent is to avoid isolation. \\

Agents $i=1,...,N$ are endowed with two variables, opinion $S_i$ and charisma $C_i$; their values are fixed. The opinions are randomly selected from a Gaussian distribution with zero mean and variance equal to 1. The charismas are randomly selected from a Poissonian distribution with mean $<C>$=3. The group size $N$ is fixed in time; once per $K=50N^2$ of time steps, one agent leaves and another agent enters; as in a queue, the leaving order is the same as the order of they entered. At each time step, each agent $i$ present in the group has a chance to express her/his opinion. Once expressed, this opinion is detected by all other agents $j=1,...,i-1,i+1,...,N$ present in the group, and it enters to the individually calculated means $<S_j>$ with the weight $C_i$. The mean value $<S_j>$, normalized by the sum of the charismas $C_i$ of the authors of expressed opinions, is perceived by $j$ as the public opinion. We assume that an opinion expressed by agent $i$ does not enter to his mean $<S_i>$. \\

Having a chance to speak, an agent $i$ expresses his/her opinion $S_i$ more likely if it is not too far from $<S_i>$. Here the probability of speaking out is calculated as

\begin{equation}
 P_i=\frac{c}{1+exp(ax_i+b)}
\end{equation}
where $x_i=\vert S_i-<S_i>\vert$
and $b,c$ are constants selected as to get speaking out unlikely for $x > 0.4$; yet, only about 20 percent of the remaining chances is used. This is obtained for $a=4.0, b=0.5, c=0.2$.\\

We are going to evaluate, how long an agent with large charisma is remembered. A natural unit of time is the lifetime of an agent in the system. Our method is akin to the damage spreading method \cite{dasp}, used in the cellular automata. Namely, we wait for an agent with large charisma ($C_j$), let us call him a leader, and we record the time dependence of the mean perceived opinion 

\begin{equation}
 <S>=\frac{\sum_{j=1}^N S_j}{N}
\end{equation}
two times, for two different signs of the opinion of the leader. As long as the difference between these two plots is non-zero, the leader is remembered.\\

Another issue here is to look for the so-called pendulum effect. The term refers to a change of trend in economy \cite{cru} or politics \cite{maa}. Here it is understood as a change of sign of the difference of the two above trajectories. \\

The results obtained within this scheme are to be repeated with two modifications. First is that the opinion, a number till now, is defined as a point in a plane. In this case, the quantity $x$ is redefined as a distance between two points quantity in a plane. Also, when presenting the results, we have to show the projection of these points on an arbitrary chosen axis. This modification is by no means a signal that we intend to enter to a difficult problem how many aspects of an issue is important for us. We just imagine the plot $<S>(t)$ as a trajectory in a space of opinions is similar to a random walk in a disordered medium, where the walker jumps from one agent's opinion to another. The point is that basically, the properties of the random walk do depend on the system dimensionality. \\

The second alternative modification is to put $P_i=c$ in Eq.1. This is simply to check how the numerical results depend on the key assumption of SST. What if not?

%% ###########################################################################
\section{Numerical results}
\label{S4}
%% ###########################################################################
 
All histograms presented here are performed with statistics of $12\times 10^3$ trajectories.\\

\begin{figure}
 \includegraphics[angle=270,width=0.86\columnwidth]{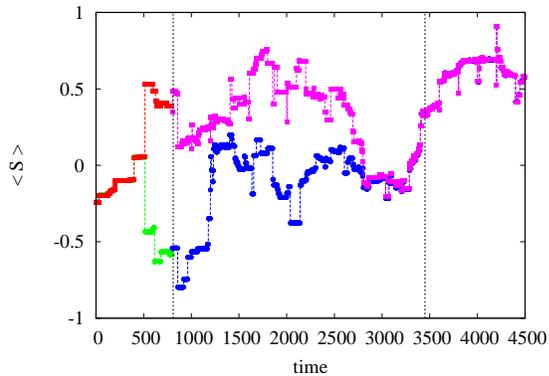} 
\caption{The time dependence of the mean opinion $<S>$ - a typical plot, where the leader's influence persists long after his leave. In this example, the leader appeared in the system at time $t=0$, and he expressed his opinion for the first time about $t=500$. The left vertical line marks the time $\tau_0$ when the leader leaves, and the right vertical line - the time $\tau$ when he is forgotten.} 
\label{fig1}
\end{figure}

\begin{figure}
 \includegraphics[angle=270,width=0.86\columnwidth]{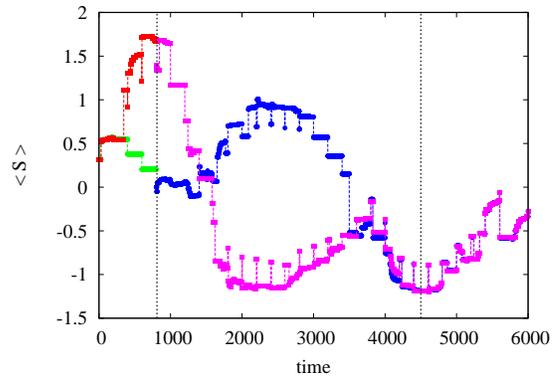} 
\caption{The time dependence of the mean opinion $<S>$ - a more rare case, where the leader's influence changes the sign, but it remains relatively large. In this example, the change of sign appears near $t=1600$. } 
\label{fig2}
\end{figure}

In Figs. \ref{fig1} and \ref{fig2} we show two exemplary plots of the mean opinion $<S>$ against time. In both cases we show two curves, driven by the same random numbers; the difference between the curves is that the leader's opinion is either positive or negative. The plot in Fig.  \ref{fig1}  is rather typical. In Fig. \ref{fig2}  we show a rare case where after the leader's leave, his influence is entirely reversed.\\

\begin{figure}
 \includegraphics[angle=270,width=0.86\columnwidth]{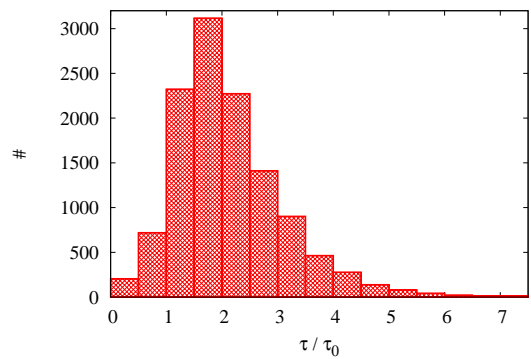} 
\caption{The histogram of the ratio $\tau/\tau_0$ for $N$=4. The results show that the time $\tau$ when a leader is remembered is in most cases at least twice longer than the leader lifetime in the system.  The results for $N=10$ are almost identical.} 
\label{fig3}
\end{figure} 

\begin{figure}
 \includegraphics[angle=270,width=0.86\columnwidth]{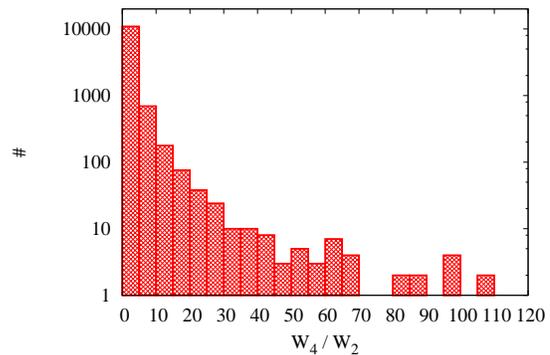} 
\caption{The histogram of the ratio $W_4/W_2$ for $N$=4. This ratio is a measure of an influence of a leader after his leave, compared with (divided by) his influence during his lifetime. The results show that quite often, the former exceeds the latter; the result exceeds the cause.} 
\label{fig4}
\end{figure} 

\begin{figure}
 \includegraphics[angle=270,width=0.86\columnwidth]{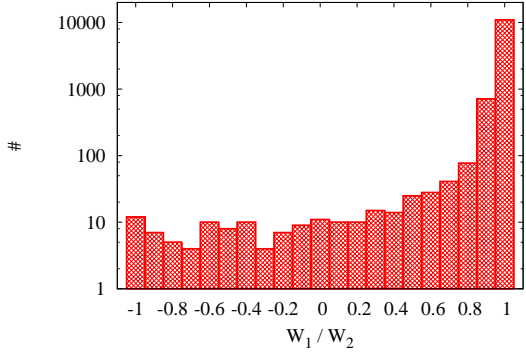} 
\caption{The histogram of the ratio $W_1/W_2$ for $N$=4. This ratio is a measure of the relevance of the pendulum effect in the presence of the leader. The results indicate that although the pendulum effect is quite frequent (the ratio is less than one), its influence remains rather small.} 
\label{fig5}
\end{figure} 

\begin{figure}
 \includegraphics[angle=270,width=0.86\columnwidth]{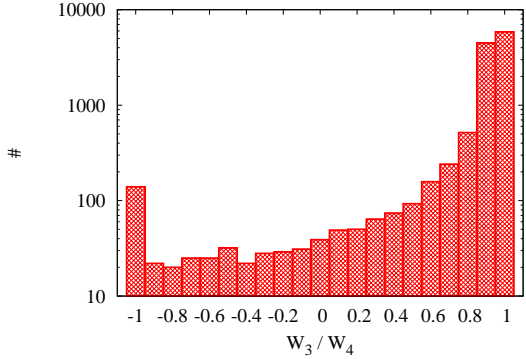} 
\caption{The histogram of the ratio $W_3/W_4$ for $N$=4. This ratio is a measure of the relevance of the pendulum effect after the leader's leave. The results indicate that the pendulum effect is slightly more frequent, than the one in the presence of the leader, shown in the previous picture \ref{fig5}. } 
\label{fig6}
\end{figure} 

\begin{figure}
 \includegraphics[angle=270,width=0.86\columnwidth]{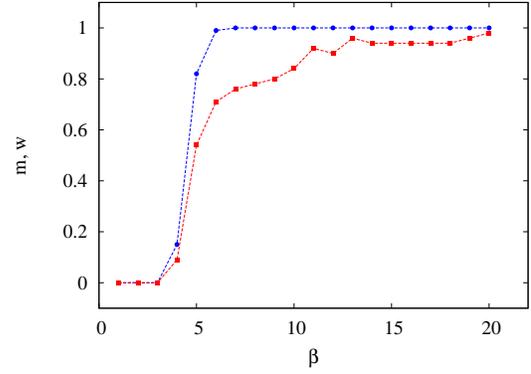} 
\caption{The percentage $w$ of program runs when, after $10^3$ iterations, the final opinion $S_{t=1000}$ is $\pm 1$ (blue circles, color online) and the value of this final opinion $m=S_{t=1000}$ averaged over 100 runs (red squares, lower curve), as dependent on the parameter $\beta$. Note that we do not use the brackets, because there is no averaging over agents; the system is equivalent to one agent who expresses random but correlated opinions.} 
\label{fig7}
\end{figure} 

Denoting the length of the time period when each agent is present in the system by $\tau _0$, we investigate the time length $\tau$ when, as told above, the two curves do not collapse into one yet. In Figs. 3 we show histograms of the $\tau/\tau_0$ ratio for $N$=4. We see that in most cases, this ratio is at least two.\\

To investigate the pendulum effect, we have to identify the cases when an influence of a leader with negative opinion exceeds the one of positive opinion. For this purpose we investigate numerically four integrals $W_i$, defined as follows:

\begin{equation}
 W_1=\int_0^{\tau_0}(<S^+>-<S^->)dt
\end{equation}

\begin{equation}
 W_2=\int_0^{\tau_0}|<S^+>-<S^->|dt
\end{equation}

\begin{equation}
 W_3=\int_{\tau_0}^{\infty}(<S^+>-<S^->)dt
\end{equation}

\begin{equation}
 W_4=\int_{\tau_0}^{\infty}|<S^+>-<S^->|dt
\end{equation}
where $S^{\pm}$ is the mean perceived opinion when the leader's opinion is $\pm 1$. The integrals $W_1$ and $W_2$ are introduced to evaluate the influence of the leader when he or she is present in the system, while $W_3$ and $W_4$ are the same measures after the leader has left. The integrals $W_1$ and $W_3$ are sensitive to the pendulum effect, while $W_2$ and $W_4$ are not.\\
  
The ratio $W_4/W_2$ is a measure of the ratio of effect to cause; the cause $W_2$ is the leader's influence when he/she is present in the group, while $W_4$ is the same influence after the leader left. As we see in Fig. \ref{fig4}, this ratio can exceed some tens, but in most cases it is close to one. This is a confirmation that the leader's influence after he has left the group, although persists for a long time, yet remains moderate.\\

We observe that when the leader is present, the rule $S^+ > S^-$ is rarely broken. Once it is broken, $W_1<W_2$; as we see in Fig. \ref{fig5}, this happens in about 15 percent of cases. Yet, when the leader leaves, the effect happens slightly more often, as shown in Fig. \ref{fig6}. One could wonder how this result depends on the system dimensionality, because in one-dimensional space the trajectories $S^+$ and $S^-$ must cross at the same point, whilst in a plane they can avoid each other. (To imagine this, we have to use time as the third axis.) However, according to our numerical results, the obtained histograms (not shown here) are very similar for one- and two-dimensional space of opinions. Also, all these effects are qualitatively the same for $N$=4 and $N$=10. It is only that large values of the $W_4/W_2$ ratio are more likely for $N$=10.\\

On the contrary, when we put $P_i=c$ in Eq.1, the results are entirely different. With the same statistics, the ratio $\tau/\tau_0$ never exceeds 1.0, and the ratio $W_4/W_2$ never exceeds 1.35. This means, that the dependence of the probability of speaking out on the perceived opinion of others is indeed the origin of the observed results.

%% ###########################################################################
\section{Is it a spiral? }
\label{S3}
%% ###########################################################################

We can ask, if the effect simulated in this way can be rightly termed 'spiral of silence'. In our view, the term is justified if the system dynamics shows some internal tendency to increase the prevalence of what is perceived as a majority opinion. To demonstrate that it is really true, let us transform the problem to a much simpler form, where only two opinions are possible, $S=\pm 1$. Also, the related probabilities of expression of these opinions are respectively $p_{t\pm}=1/2(1 \pm rS_t)$; this is justified for small $S_t$. Let the perceived opinion be updated in discrete time steps as 

\begin{equation}
 S_t \to S_{t+1} =\alpha S + (1-\alpha)S_t
\end{equation}
where $S_t$ is the opinion perceived at time $t$, $S$ is the opinion expressed by one of agents at time, say, between $t$ and $t+1$ and $\alpha <1$ is a weight constant. In the 
average we get

\begin{eqnarray*}
S_{t+1}&=&p_{t+}[\alpha+(1-\alpha)S_t]+p_{t-}[-\alpha+(1-\alpha)S_t]\\
&=&(r\alpha+1-\alpha)S_t		\\(4)
\end{eqnarray*}
In particular, if the perceived opinion at time $t$ is $\varepsilon$, the mean opinion perceived at time $t+1$ is $(r\alpha+1-\alpha)\varepsilon$. The result is larger than $\varepsilon$ iff $r>1$, irrespectively on the weight $\alpha$. Varying $r$, we get the classical pitchfork bifurcation \cite{glen}; for $r>1$, the fixed point $S_t=0$ is unstable. This means, that an opinion perceived as a majority opinion displays a tendency to self-establish, even if initially it is small. In this way the term 'spiral' seems justified here. We note that to reach this, the concept of a threshold is not needed, on the contrary to all approaches cited above. \\

However, we should add that the calculation presented in this section relies on a rough procedure of averaging over two possible operations; if $S_t=+1$, the perceived positive opinion is enhanced, while for $S_t=-1$ it goes down. A more accurate analysis could be done in terms of correlated iterative equations, with the correlation changing dynamically with the perceived opinion through the probabilities $p_{t\pm}$. We note that the linear transformations (7) for two different signs of $S$ do not commute. We have performed a set of transformations, with $S_0=0.1$ as a starting point, and $\alpha=0.05$. The correlations between subsequent signs of $S$ are introduced via the probability  $(1+\exp(\beta))^{-1}$ that an expressed opinion is different than the previously expressed one. In Fig. \ref{fig7} we show how the asymptotic value of $<S>$ depends on the coefficient $\beta$. The results indicate that once the coefficient $\beta$ is larger than five, the system stabilizes at $S_t=\pm 1$. 

%% ###########################################################################
\section{Discussion}
\label{S5}
%% ###########################################################################

The model presented here contains some ad-hoc assumptions which have not been relaxed. One of them is a specific relation of $K(N)=50N^2$. There are other functions of type of $K(N)\propto N^\alpha$, which seem equally reasonable. We do not expect qualitative changes in the result, if the proportionality constant is of order of tens. Similarly, the form of the function $P_i$ in Eq. (1) is arbitrary, as well as the constants $a,b,c$ therein. Yet we believe that these functions play their role: they allow to capture the investigated effect without wasting too much of computational time.\\
 
New elements of this work are as follows. First, the model of the effect of spiral of silence is reformulated without reference to the concept of threshold. The latter concept is vivid in previous formulations of TSS: \cite{gso,krss} and their followers. Second new element is the dynamic frame of our model, namely the idea of an open system. Without this modification, the model outcome is expected to be reduced to an equilibrium, with the leader's influence frozen-in the final state. We consider such a result to be artificial; actually, our model provides a natural mechanism of gradual vanishing the memory on the initial state. Third, this novelty allows to evaluate in a model way the time of this forgetting process. Our main result is that it can take some generations of agents. This persistence of memory concides with the result, that the pendulum effect is relatively rare. \\

If we take the model results directly, we could conclude that the memory of a leader persists, transmitted to agents who have seen neither the leader, nor his/hers direct successors. As it was pointed by Bourdieu \cite{bou} the determination of action could be driven by usage, tradition or custom leading to formulation of habitus. It seems to us surprising that this process can be driven by the spiral of silence. Yet, continuing for a while the analogy offerred by the model, we can imagine that if everybody can say anything, it is hard to distinguish what is to be remembered, unless we have to deal with orchestration of habitus which allowed us mutual anticipation of behavior of others according to a common sense approach -- I know that you know that I know.\\

%% ###########################################################################
\begin{acknowledgments} 
The work was partially supported by the Polish Ministry of Science and Higher Education and its grants for Scientific Research, and by the PL-Grid Infrastructure.
\end{acknowledgments}
%% ###########################################################################

 \end{document}